# A Secure Communication in Mobile Agent System

Maryam Mahmoodi [#1], Mohammad Mahmoodi Varnamkhasti [#2]

[1]*Islamic Azad University, Meymeh branch, Department of computer, Meymeh, Iran*
[2]*Department of computer, University of Isfahan*

*Abstract*—**A mobile agent is a software code with mobility which can be move from a computer into another computers through network. The mobile agent paradigm provides many benefits in developments of distributed application at the same time introduce new requirements for security issues with these systems. In this article we present a solution for protection agent from other agents attacks with loging patterns of malicious agent and useing this log for communication. We implemented our resolution by JADE.**

*Keywords—*mobile agent; security; mobility; agent security

## I. INTRODUCTION

A mobile agent is a software program with mobility which can be sent out from a computer into a network and roam among the computer nodes in the network[3]. It can be executed on those computers to finish its task on behalf of it's owner. This mobility allow mobile agent to travel between one or more remote Computer. The key characteristic of the mobile agent paradigm is that any host in the network is allowed a high degree of flexibility to possess any mixture of skills, resources and processors. In mobile agents, the mobile code generated by one party transfers and execute in an environment controlled by another party so several security issues arise in various mobile agent computing. These issues include authentication, authorization (or access control), intrusion detection etc. Because of mobility of mobile agent, the security problems becomes more complicated and have become a bottleneck for development and maintenance of mobile agent technology especially in security sensitive applications such as e-commerce, military applications, scientific applications etc[4]. In this article we explain various security objectives and requirements, the in section 2 we discusses various security threats and attacks on mobile agents. Section 3 explains security mechanisms and section 4 includes our solution for protect agent from agent attacks.

## II. SECURITY REQUIREMENTS

### A. Authentication and authorization

Authentication is the process of verifying the identity or other relevant information about the entity. The outcome of the authentication processes is that the user/ agent knows the identity of the server/agent execution environment and server/agent execution environment knows the identity of the user/agent. The process of deciding whether or not to grant a request after confirmation about the authentication of the principal is called authorization or access[12]. In this case digital signatures are required in addition to password access.

### B. Accountability and non-Repudiation

This objective requires that users and administrators accept themselves acts and requires that either side of a communication cannot deny the communication later. For this we should log important communication exchanges to prevent later denials. We need to record not only unique identification and authentication but also an audit log of security relevant events to which both agent or process responsible for those events and security related activities must be recorded in this log[1] .

### C. Privacy, confidentiality and anonymity

Privacy and confidentiality include confidentiality of exchanges and interactions in a mobile agent system. In anonymity platform should keep agent's identity when necessary and legal.

### D. Availability

This objective include ensures availability of both data and services of a mobile agent to local agents and remote agents. This agent platform should ensure availability of controlled concurrency, support for simulation access, deadlock management and exclusive access when required[2].

### E. Fairness

No part can give advantage over other parts. So, mechanisms are necessary to ensure fair agent platform interaction in electronic exchange.

### F. Integrity

The agent platform must protect agents from unauthorized modification of their code, state and data and ensure that only authorized agents or processes carry out any modification of shared data.

### III. SECURITY THREATS

Threats to security generally fall into three main classes: disclosure of information, denial
of service, and corruption of information. System include two main components:

Agent and agent platform. Here an agent is comprised of the code and state information needed to carry out some computation. Mobility allows an agent to move or hop among agent platforms. Generally threats include:

#### A. Masquerading

When an unauthorized agent claims the identity of another agent it is said to be
masquerading. The masquerading agent may pose as an authorized agent in an effort to
gain access to services and resources to which it is not entitled. A masquerading agent platform can access to information of agents that exist in itself and share them to another platforms.

#### B. Denial of Service

Mobile agents can launch denial of service attacks by consuming an excessive amount of
the agent platform's computing resources. These attacks can be launched intentionally by running attack scripts to exploit system vulnerabilities, or unintentionally through programming errors.[2] Program testing, configuration management, design reviews, independent testing, and other software engineering practices have been proposed to help reduce the risk of errors.
This attacks may happen by platform to agent. When an agent arrives at an agent platform, it expects the platform to execute the agent's requests faithfully. A malicious agent platform, may ignore agent service requests, introduce unacceptable delays for critical tasks.

#### C. Unauthorized Access

Access control mechanisms are used to prevent unauthorized users or processes from accessing services and resources for which they have not been granted permission and privileges as specified by a security policy. Each agent visiting a platform must be subject to the platform's security policy. A platform must ensure that agents do not have read or write access to data for which they have no authorization.

#### D. Repudiation

Repudiation occurs when an agent, participating in a transaction or communication, later
claims that the transaction or communication never took place. So must the agents and agent platforms involved in the transaction maintain records to help resolve any dispute.

#### E. Eavesdropping

It involves the interception and monitoring of secret communications. The threat of eavesdropping, however, is further exacerbated in mobile agent systems because the agent platform can not only monitor communications, but also can monitor every instruction executed by the agent, all the unencrypted or public data it brings to the platform, and all the subsequent data generated on the platform.

Generally, 4 threats categories are identified:
-Agent- to-Platform threats
-Agent-to-Agent threats
-Platform-to-Agent threats
-Other-to-platform threats

Agent to platform threats include masquerading denial of service and unauthorized access. Agent to agent threats include masquerade, denial of service, repudiation and unauthorized access. Platform to agent threats include masquerade, denial of service, eavesdropping and alteration and other to agent platform threats include masquerade, unauthorized access and denial of service.

### IV. SECURITY MECHANISMS

This Security mechanisms provide assurance that remote hosts will adhere to policies for Mobile Code programs[1]. These mechanisms can detect attacks or prevent them[5,6].

#### A. Detection mechanisms

These mechanisms enable a mobile agent's owner to identify that an attack has occurred on the mobile code program. This helps in analyzing validity of results that program has accumulated but only when attack has done its work[8].

#### B. Prevention mechanisms

The mechanisms here try to make it impossible or very difficult to access or modify code, state or execution flow of a mobile agent[7].

### V. OUR SOLUTION

Our solution is about agent protection. It protects agents against malicious agents and platforms attacks. Execution tracing[9] is one of the best technique in this matter, that is a technique for detecting unauthorized modifications of an agent through the recording of the agent's behavior during it's execution on each agent platform. The technique requires each platform involved to create and retain a non-repudiatable log or trace of the operations performed by agent while resident there and to submit a cryptographic hash of the trace upon conclusion as a trace summery or fingerprint. A trace is composed of a sequence of statement identifiers and platform signature information. The technique also defines a secure protocol to convey agents and associated security related information among the various parties involved. If any suspicious results occur, the appropriate traces and trace summaries can be obtains and verified and malicious host identified.

But it's obvious that the size and number of logs increase continually and increase overhead. In our solution, we let system to works and detects malicious agent and platform and retain them in log instead of retain execution trace information.

Since requests of malicious agents are special binary string, system can detect pattern of this agent and retain them in log and use them for future communications. In execution and gather this information about malicious agents, we can protect agent's code, state and state by cryptographic protocols[10,14,13,12] and encapsulation methods. Existing cryptographic protocols can be easily embedded in our solution.

We have implemented our solution with JADE which is a distributed environment java based. Each JADE agent platform can be split into distributed agent containers and inter_cotainer mobility mechanisms are already in place, We embedded the log file and the cryptographic services at each agent.

## VI. CONCLUSION

A mobile agent is a software code with mobility which can be move from a computer into another computers through network. So, we must understand and develop security mechanisms that both detect and prevent malicious attacks against mobile code program. In this article we explain security requirements, security objectives and security mechanisms but most of the security measures are n't adequate for example about agent to agent attacks can log execution trace but size of the log is a problem.

In this paper we present a solution and it is log patterns of malicious agents and embedded this at agents. Agent use it for detect authorized agents and it's communications. We have implemented our solution with JADE.

## REFERENCES

[1] P. Dadhich, K.Dutta, M.C.Govil, "Security Issues in Mobile Agents", International Journal of Computer Applications (0975 – 8887)Volume 11– No.4, December 2010.

[2] W.Jansen, T.Karygiannis, "Mobile Agent Security" , NIST Special Publication 800-19,

[3] Colin G. Harrison, David M. Chess, and Aaron Kershenbaum, "Mobile Agents: Are they a good idea?", technical report, 1995, IBM Research Division.

[4] Chess, D.M: Security issues in mobile code systems. In : mobile agents and security, Editor Vigna, vol. LNCS1419. Springer-Verlag 1998.

[5] Ahmed S. Mohamed & D. Fakhry, Security in Mobile Agent Systems. In proceedings of the 2002 Symposium on Applications and the Internet, pgs 4-5, Washington, DC, USA,2002, IEEE Computer Society.

[6] B. Yee. Using Secure Coprocessors . Ph.D thesis , Carnegie Mellon University. 1994.

[7] U. G. Wilhelm," A Technical Approach to Privacy based on mobile Agents Protected by Tamper-resistant Hardware", PhD Theses nr. 1961. Dept. of D'Informatique, Ecole polytechnique Federale de Lausanne,1999.

[8] J.Algesheimer et al.,"Cryptographic Security for Mobile Code,"Proc.2001 IEEE Symp. Security andPrivacy, IEEE Press,2001

[9] Giovanni Vigna, "Protecting Mobile Agents Through Tracing," Proceedings of the 3rd ECOOP Workshop on Mobile Object Systems, Jyväskylä, Finland, June 1997.

[10] Neeran M. Karnik and Anand R. Tripathi. Security in the Ajanta mobile agent system. Software Practice and Experience, 31(4):301–329, 2001.

[11] Chess David M. Security issues in mobile code systems. In Mobile Agents and Security, volume 1419, pages 1–14. Springer Verlag, 1998.

[12] J. Mir and J. Borrell. Protecting mobile agent itineraries. In Mobile Agents for Telecommunication Applications (MATA), volume 2881 of Lecture Notes in Computer Science, pages 275–285. Springer Verlag, October 2003.

[13] S. Robles, J. Mir, and J. Borrell. Marism-a: An architecture for mobile agents with recursive itinerary and secure migration. In 2nd. IW on Security of Mobile Multiagent Systems, Bologna, July 2002.

[14] V. Roth. Empowering mobile software agents. In Proc. 6th IEEE Mobile Agents Conference, volume 2535 of Lecture Notes in Computer Science, pages 47–63. Spinger Verlag, 2002.